\documentclass[preprint, pdf]{iucr}              
\RequirePackage{graphicx}
     \journalcode{S}              

\begin{document}                  

\title{Water Window Ptychographic Imaging with Characterized Coherent X-rays}


\author[a]{Max}{Rose}
\author[a,b]{Petr}{Skopintsev}
\author[a,c]{Dmitry}{Dzhigaev}
\author[a,d]{Oleg}{Gorobtsov}
\author[e,f,g]{Tobias}{Senkbeil}
\author[e,f,g]{Andreas}{von Gundlach}
\author[e,f,g]{Thomas}{Gorniak}
\author[a]{Anatoly}{Shabalin}
\author[a]{Jens}{Viefhaus}
\author[e,f,g]{Axel}{Rosenhahn} 
\cauthor[a,c]{Ivan}{Vartanyants}{ivan.vartaniants@desy.de}

\aff[a]{Deutsches Elektronen-Synchrotron DESY, Notkestrasse 85, 22607 Hamburg, \country{Germany}}
\aff[b]{Moscow Institute of Physics and Technology (State University), Dolgoprudny, Moscow Region 141700 \country{Russia}}
\aff[c]{National Research Nuclear University 'MEPhI' (Moscow Engineering Physics Institute), Kashirskoe shosse 31, 115409 Moscow, \country{Russia}}
\aff[d]{National Research Center, 'Kurchatov Institute', Kurchatov Square 1, 123182 Moscow, \country{Russia}}
\aff[e]{Analytical Chemistry - Biointerfaces, Ruhr-University Bochum, Universitätsstraße 150, 44780 Bochum, \country{Germany}}
\aff[f]{Applied Physical Chemistry, University of Heidelberg, Im Neuenheimer Feld 253, 69120 Heidelberg, \country{Germany}}
\aff[g]{Institute of Functional Interfaces, Karlsruhe Institute of Technology, Eggenstein-Leopoldshafen, \country{Germany}}

\maketitle                        

\begin{abstract}
We report on a ptychographical coherent diffractive imaging experiment in the water window with focused soft X-rays at $500~\mathrm{eV}$.
An X-ray beam with high degree of coherence was selected for ptychography at the P04 beamline of the PETRA III synchrotron radiation source.
We measured the beam coherence with the newly developed non-redundant array method.
A pinhole $2.6~\mathrm{\mu m}$ in size selected the coherent part of the beam and was used for ptychographic measurements of a lithographically manufactured test sample and fossil diatom.
The achieved resolution was $53~\mathrm{nm}$ for the test sample and only limited by the size of the detector.
The diatom was imaged at a resolution better than $90~\mathrm{nm}$.
\end{abstract}



\section{Introduction}

Imaging in the water window energy range between the absorption edges of carbon and oxygen at
$284~\mathrm{eV}$
and
$532~\mathrm{eV}$
yields a high chemical contrast in objects mainly composed of carbon and its aqueous components
\cite{larabell2010a}.
The coherent X-ray diffractive imaging (CXDI) method
\cite{miao1999a}
applied with third-generation synchrotron sources has proven to be a useful tool in structural analysis on the nanoscale
\cite{chapman2010b,mancuso2010a,murphy2013}.
In CXDI no lenses are used and this in principle allows to overcome the resolution limitations of conventional lens microscopes. Reconstruction of an object from measured diffraction patterns requires solving the well-known phase problem. Iterative phase retrieval techniques have been successfully employed to solve the phase problem
\cite{fienup1982a,Marchesini2007a}.
As a limitation, the CXDI technique requires the sample to be isolated and fully illuminated by the coherent X-ray beam. In order to study extended objects with X-rays and to improve the uniqueness and convergence of the phase retrieval process, ptychographic coherent diffractive imaging (PCDI) was employed
\cite{rodenburg2007a}.
PCDI involves scanning of the X-ray beam along the object up to a desired field of view. A certain overlap of the illuminated areas is crucial to succeed with the phase retrieval and the image reconstruction
\cite{bunk2008a}.
Conventional CXDI and earlier iterative algorithms for PCDI required precise knowledge of the probe for example the X-ray beam intensity profile incident on the object. \emph{A priori} knowledge of the probe in ptychography is no longer necessary with algorithms that retrieve object and probe simultaneously
\cite{thibault2008a}.
Moreover, ptychography has become an excellent tool to characterize optical elements such as pinholes
\cite{giewekemeyer2010a},
zone plates
\cite{thibault2008a},
focusing mirrors
\cite{kewish2010a},
compound refractive lenses
\cite{schropp2010a},
X-ray waveguides
\cite{giewekemeyer2010b}
and effects on the phase of the wave field in the focus
\cite{dzhigaev2014a}.

Third generation synchrotron sources provide intense and highly coherent X-rays which are widely used for coherent diffractive imaging on the nanoscale. The degree of coherence may not be constant along the cross section of the X-ray beam
\cite{vartanyants2003a}.
We determined the coherence properties of the X-ray beam before performing our PCDI experiment and selected the most coherent part of the beam to avoid degradation of contrast in the diffraction patterns. Partial coherence may cause artifacts in the image reconstructions
\cite{vartanyants2001a}
and limits the spatial resolution. The most direct strategy to determine the spatial coherence of soft X-rays from synchrotron sources is to perform a set of Young’s double slit experiments with different separation between the slits
\cite{chang2000a,paterson2001a}.
At the same time the approach of coherence characterization by non-redundant arrays (NRA) recently applied to X-rays
\cite{skopintsev2014a}
offers a fast and reliable method to obtain the spatial coherence properties of synchrotron radiation from a single measurement.

Fossil diatoms are unique monads with a light weight exoskeleton consisting of silicon dioxide (SiO$_2$). They exhibit very fine periodic three dimensional structures on the nano- and microscale at the same time which can hardly be manufactured by current nano-technology methods. A fossil diatom can thus be used as X-ray resolution test sample made by nature. Previously silica shells of fossil diatoms were successfully studied with coherent imaging methods at synchrotron
\cite{giewekemeyer2010a}
and free-electron laser sources
\cite{mancuso2010b}.
The diatom investigated in this work belongs to the dominating species of nano-planktonic pennate \emph{Fragilariopsis cylindrus} that is typically found in ice-edge zones in Antarctic waters
\cite{kang1992a}.

In this paper, we first present the characterization of the spatial coherence of the soft X-ray beamline with an NRA. This is followed by two PCDI measurements in the water window with optimized beam coherence. A lithographically manufactured test pattern of known structure and the fossil diatom are reconstructed as high resolution amplitude and phase contrast projection images.

\section{Experiment}

The experiment was performed at the soft X-ray beamline P04
\cite{Viefhaus2013a}
at the PETRA III synchrotron radiation facility at DESY in Hamburg. The schematic layout of the beamline is shown in
Figure 1(a).
An APPLE-II type helical undulator of 5 m length with 72 magnetic periods was tuned to deliver photons at an energy of 500 eV which corresponds to a wavelength of
$\lambda = 2.5$ nm.
The beam propagated to the dedicated X-ray vacuum scattering chamber
[Holografische Roentgen-Streuapparatur, HORST,
\cite{Gorniak2014a}]
through several optical elements, including a beam-defining slit (27 m downstream from the undulator), a horizontal plane mirror (35 m) and a monochromator unit consisting of a vertical plane mirror together with a plane varied-line-spacing (VLS) grating (46 m). The VLS grating focused the beam at the exit slit (71 m). A cylindrical mirror (79.1 m) collimated the beam in horizontal direction (not shown in Figure 1(a)). An elliptical mirror (78.5 m) focused the beam in vertical direction to the sample position (81 m). All mirrors were designed to accept a root mean square (rms) beam size of
$6\sigma$.

Knowledge about the coherence properties is an important prerequisite for CXDI experiments. For our ptychographic measurements we defined the size of the probe incident on the sample and selected the most coherent part of the X-rays at the same time with a pinhole of
2.6 $\mu$m
diameter that was etched with a focused ion beam in a
2 $\mu$m
gold layer on a
$100$ nm
thin
$\mathrm{Si_{3}N_{4}}$
membrane.

\begin{figure}
\centering
\includegraphics[width=\linewidth]{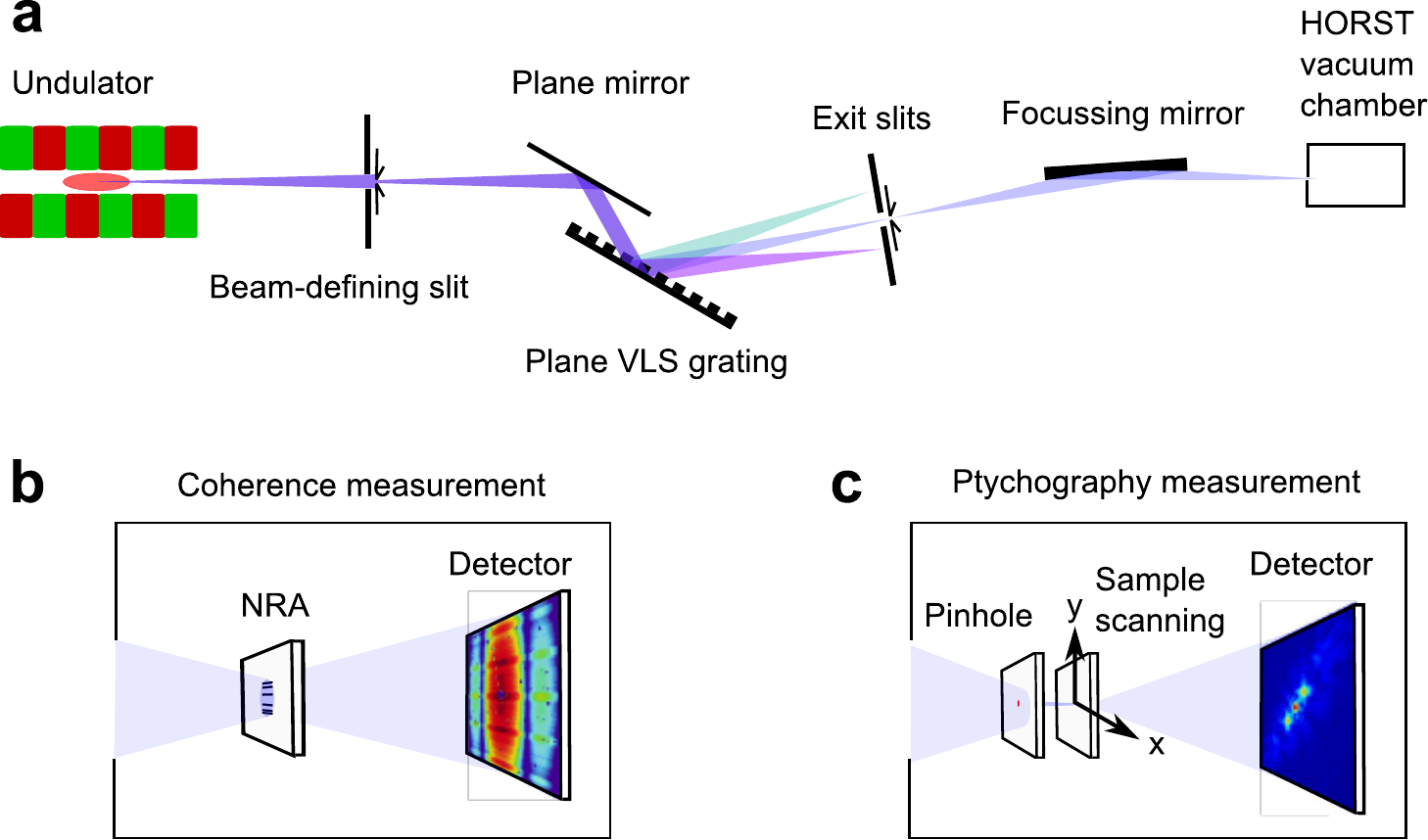}
\caption{(a) Soft X-ray beamline layout. The X-ray scattering vacuum chamber HORST in (b) Coherence measurement setup and (c) ptychography setup.}
\label{fig:setup}
\end{figure}

In both cases of the coherence and ptychography measurements the same sample holder of the HORST chamber was used
(see Figure \ref{fig:setup}(b) and (c)).
It consisted of a closed-loop piezo-electric stage to allow horizontal and vertical scanning of the sample relative to the probing X-ray beam with accuracy below
$20$ nm.
The far-field diffraction pattern intensities were measured by a CCD detector
(DODX436-BN, Andor Technology Ltd., Belfast, UK).
The square detector area of
27.6 x 27.6 mm\textsuperscript{2}
consisted of
2048 x 2048 pixels
with a pixel size of
13.5 x 13.5  $\mathrm{\mu m}^2$.

\section{Coherence Characterization}

\subsection{Theory}

A brief description of coherence theory is given in the following section to explain how the coherence properties of the beam can be retrieved from a single NRA diffraction pattern
\cite{skopintsev2014a}.
In the theory of optical coherence, the statistical properties of the radiation are described by the mutual coherence function (MCF)
$\Gamma_{12}(\tau)$
\cite{Goodman2000a, Mandel1995a}

\begin{equation}
\Gamma_{12}(\tau) =
\langle
E^*(\mathbf{r}_1 , t) \cdot E(\mathbf{r}_2 , t + \tau)
\rangle
,
\label{eq:MCF}
\end{equation}

where
$E(\mathbf{r}_1 , t)$
and
$E(\mathbf{r}_2 , t + \tau)$
are the field values at positions and times
$\mathbf{r}_1 , t$
and
$\mathbf{r}_2 , t + \tau$,
and the brackets
$\langle \cdots \rangle$
indicate the average over time. The intensity
$I_i$
at position
$\mathbf{r}_i$
is given by
$\langle
|E(\mathbf{r}_i , t)|^2
\rangle$.
The complex degree of coherence
$\gamma_{12}(\tau)$
is defined as the normalized MCF

\begin{equation}
\gamma_{12}(\tau) =
\frac{\Gamma_{12}(\tau)}{\sqrt{
\langle
|E(\mathbf{r}_1 , t)|^2
\rangle
\cdot
\langle
|E(\mathbf{r}_2 , t)|^2
\rangle
}}
.
\label{eq:cdcoh}
\end{equation}

When the time delay
$\tau$
is much shorter than the coherence time
$\tau_{\mathrm{c}}$,
the complex degree of coherence can be approximated by the complex coherence factor (CCF)
$\gamma_{12} = \gamma_{12}(0)$
\cite{Goodman2000a}.
To characterize coherence by a single quantity the global degree of coherence
$\zeta$
is often introduced as
\cite{Vartanyants2010a}

\begin{equation}
\zeta =
\frac{\int{
|\gamma_{12}|^2 I(\mathbf{r}_1) I(\mathbf{r}_2) \, \mathrm{d}\mathbf{r}_1 \, \mathrm{d}\mathbf{r}_2 }}
{\int{
I(\mathbf{r}_1) \, \mathrm{d}\mathbf{r}_1}
\cdot
\int{
I(\mathbf{r}_2) \, \mathrm{d}\mathbf{r}_2
}}
.
\label{eq:gdcoh}
\end{equation}

In the frame of the Gaussian Schell-model (GSM), which in most cases provides sufficient physical description of the synchrotron radiation, the intensity profile and the CCF are both considered to be Gaussian functions
\cite{Mandel1995a}.
In this model the partially coherent beam is characterized by the standard deviation
$\sigma$
of the beam size and its transverse coherence length
$l_{\mathrm{coh}}$.
The coherence length is defined as the standard deviation of the modulus of the CCF
$|\gamma_{12}|$.
In the frame of GSM, the global degree of coherence
$\zeta$
from equation (\ref{eq:gdcoh}) can be expressed as
\cite{Vartanyants2010a}

\begin{equation}
\zeta = \left( l_{\mathrm{coh}}/\sigma \right)
 \left[ 4+ \left( l_{\mathrm{coh}}/\sigma \right) ^2 \right]^{-1/2} \ .
\label{eq:gdcohsigma}
\end{equation}

A Non-redundant array of apertures can be used to measure the CCF. It is shown
\cite{Mejia2007a,skopintsev2014a}
that for narrow bandwidth radiation the intensity
$I(\mathbf{q})$
of the far-field interference pattern as a function of the momentum transfer vector
$\mathbf{q}$
observed in a diffraction experiment with
$N$
apertures is

\begin{equation}
I(\mathbf{q}) = I_S(\mathbf{q})
\left \{
C_0 + \sum \limits_{i \neq j}^{N} C_{i,j} \left[ e^{\mathrm{i} \alpha_{i,j} \delta (\Delta \mathbf{x} - \mathbf{d}_{i,j})} \right]
\right \}
.
\label{eq:NRAintensity}
\end{equation}

Here
$I_S(\mathbf{q})$
is the diffraction pattern of a single aperture. Individual aperture separations are denoted by
$\mathbf{d}_{i,j} = -\mathbf{d}_{j,i}$
and
$\alpha_{i,j} = -\alpha_{j,i}$
are the relative phases. For the analysis of the diffraction pattern from
equation (\ref{eq:NRAintensity})
its Fourier transform is used

\begin{equation}
\hat{I}(\Delta\mathbf{x}) = \hat{I}_S(\Delta\mathbf{x}) \otimes
\left \{
C_0 \delta(\Delta\mathbf{x}) + \sum \limits_{i \neq j}^{N} C_{i,j} \left[ e^{\mathrm{i} \alpha_{i,j}} \delta (\Delta \mathbf{x} - \mathbf{d}_{i,j}) \right]
\right \}
.
\label{eq:NRAintensityFT}
\end{equation}

Here
$\delta(\mathbf{x})$
is the Dirac delta function,
$\hat{I}_S(\Delta\mathbf{x})$
is the Fourier transform of a single aperture diffraction intensity
$I_S(\mathbf{q})$
and the symbol
$\otimes$
denotes the convolution. The coefficient
$C_0$
is defined as
$C_0 = \sum \limits_{i=1}^{N} I_i$,
where
$I_i$
is the intensity incident on the
$i$-th
aperture. The coefficients
$C_{i,j}$
are equal to the mutual intensity
$\Gamma_{i,j}(0)$
at
$\tau = 0$

\begin{equation}
C_{i,j} = \Gamma_{i,j}(0) = |\gamma_{i,j}| \sqrt{I_i I_j}
.
\label{eq:Coeff}
\end{equation}

To each individual aperture separation
$\mathbf{d}_{i,j}$
corresponds a single peak in
equation (\ref{eq:NRAintensityFT})
with its height being equal to
$C_{i,j}$.
The intensities
$I_i$, $I_j$
together with peak heights
$C_{i,j}$
are used to obtain the CCF values from
equation (\ref{eq:Coeff})

\begin{equation}
|\gamma_{i,j}| = \frac{C_{i,j}} {\sqrt{I_i I_j}}
.
\label{eq:cdcohCoeff}
\end{equation}

\subsection{Results of the NRA coherence measurement}

\begin{figure}
\centering
\includegraphics[width=\linewidth]{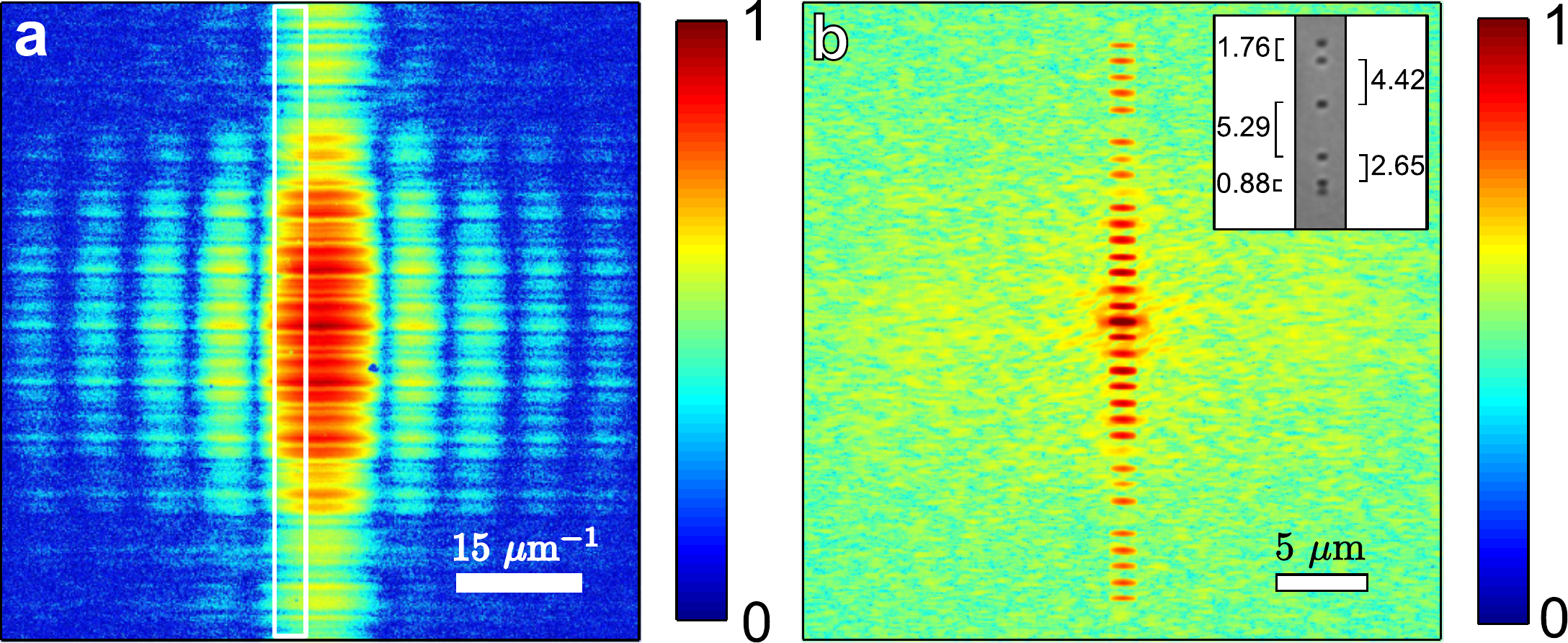}
\caption{(a) Diffraction pattern from the NRA diffraction measurement at
500 eV
and
$D_{\mathrm{es}} =$ 100 $\mu$m.
The white rectangle indicates the area used for the analysis. (b) The Fourier transform of the NRA diffraction pattern. Both images are displayed on a logarithmic scale. (Inset) Optical microscope image of the NRA and its aperture separations are shown in microns.}
\label{fig:NRA}
\end{figure}

The coherence properties of the P04 beamline were measured with a single NRA diffraction pattern for each set of beamline parameters and analyzed using the approach described in the previous section. The coherence was determined for
50 $\mu$m, 100 $\mu$m and 200 $\mu$m
monochromator exit slit opening
$D_{\mathrm{es}}$.
From the monochromator resolving power of
$E / {\Delta E} = \lambda / {\Delta \lambda} = 6 \cdot 10^3$
the temporal coherence length
$l_\mathrm{t} = \lambda \cdot \lambda / {\Delta \lambda}$
was estimated to be
15 $\mu$m.
Geometric considerations showed that the maximum optical path length difference in our experiment was
$l = 0.2$ $\mu$m
in the region used for the coherence analysis. This was much smaller than the temporal coherence length and confirmed our approximation of the complex degree of coherence by the CCF in the previous section. The spatial coherence of the X-ray beam was obtained by measuring the diffraction pattern produced by the NRA at the detector positioned
1 m
downstream
(see Figure \ref{fig:setup}(b)).
The NRA consisted of
$N = 6$
identical rectangular apertures. Each aperture was
0.8 x 0.25 $\mu$m\textsuperscript{2}
in size and manufactured according to a Golomb ruler
\cite{Lam1988a}. Our NRA was a Golomb ruler of order
6
with each separation between two individual apertures being unique
(see inset in Figure \ref{fig:NRA}(b)).
A background-corrected diffraction pattern of the NRA is shown in
Figure \ref{fig:NRA}(a).
The Fourier transform
$\hat{I}(\Delta\mathbf{x})$
of the measured diffraction pattern from the NRA is presented in
Figure \ref{fig:NRA}(b)
and shows
31
well separated peaks.

To determine the CCF we analyzed the area (51 x 2001 pixels) shown as the white rectangle in
Figure \ref{fig:NRA}(a).
In this region 51 line scans were used to determine the peak heights
$C_0$
and
$C_{i,j}$.
This area corresponds to the part of reciprocal space where the contribution of the high harmonics of X-ray radiation from the undulator is minimal
\cite{skopintsev2014a}.

\begin{figure}
\centering
\includegraphics[width=\linewidth]{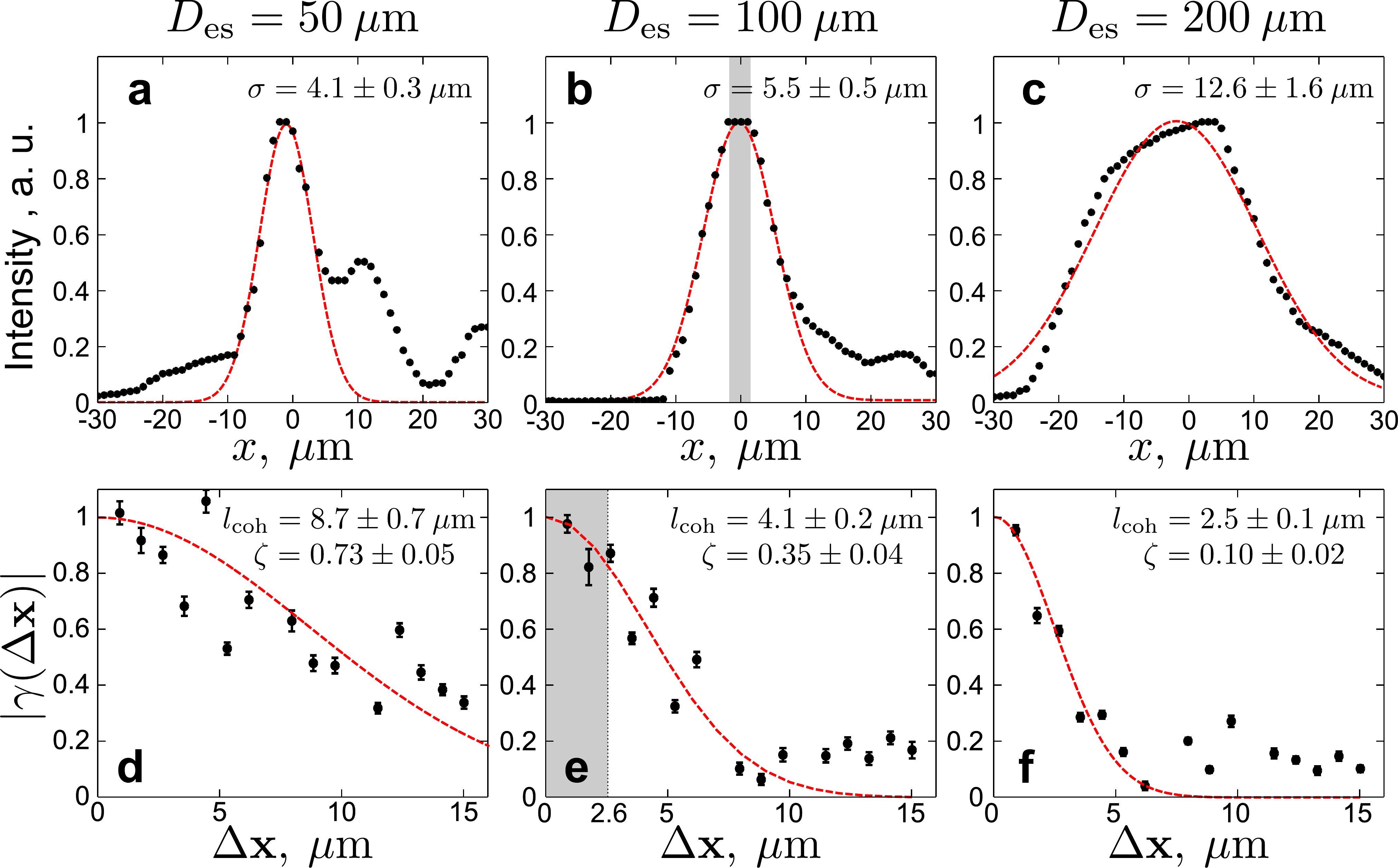}
\caption{Results of coherence measurement for three exit slit openings
$D_{\mathrm{es}}$
at
500 eV.
Black dots indicate measured data and dashed lines represent Gaussian fits. (a)-(c) Intensity profiles measured with scans of a
1.5 $\mu$m
pinhole. (d)-(f) Modulus of the CCF
$|\gamma( \Delta\mathbf{x} )|$.
The gray shaded area in (b) and (e) indicate the coherent part of the beam selected by the beam defining
2.6 $\mu$m
pinhole. The rms values
$\sigma$
of the beam size obtained from Gaussian fits, the coherence length
$l_\mathrm{coh}$
as well as the values of the global degree of coherence
$\zeta$
determined from
equation (\ref{eq:gdcohsigma}) are also shown.
For the Gaussian fits we used the data points up to
9 $\mu$m
in (e) and up to
7 $\mu$m
in (f).}
\label{fig:coherenceCurves}
\end{figure}

The intensities
$I_i$
in the focus of the beam were determined by a beam profile scan with a pinhole of
1.5 $\mu$m
diameter and are shown in
Figure \ref{fig:coherenceCurves}(a-c).
For the
50 $\mu$m
exit slit opening the intensity profile has a narrow peak. The side lobe at the right side of the profile was the result of diffraction from imperfections of the exit slit edges. At the
100 $\mu$m
exit slit opening the side lobe has almost disappeared because a different section of the exit slit edge was illuminated. At the
200 $\mu$m
exit slit opening a broad profile is observed without effects from the slit imperfections. The relative difference of the photon flux for each exit slit opening was measured by a photo diode. For the
200 $\mu$m
slit opening the highest flux was observed. In the case of
100 $\mu$m and 50 $\mu$m
the flux was reduced by a factor of
2.5 and 8,
respectively. The flux was expected to be linearly dependent on the exit slit opening. We attributed the deviation to the imperfections and uncertainty of the exit slit positioning system at exit slit openings smaller than
100 $\mu$m.

The CCF was obtained from intensities
$I_{i}$
and 51 sets of
$C_{0}$, $C_{i,j}$.
For each set of parameters
$C_{0}$, $C_{i,j}$
the modulus of the CCF as a function of the NRA aperture separation
$\Delta \mathbf{x}$
was retrieved using
equation (\ref{eq:cdcohCoeff}).
In Figure \ref{fig:coherenceCurves}(d)-(f) the averaged CCF
$|\gamma_{i,j}|$
with error bars denoting the standard deviation is presented.

The spatial coherence length
$l_\mathrm{coh}$
was obtained by the Gaussian approximation
$\exp{ \left( -\Delta\mathbf{x}^2 / 2 l_\mathrm{coh}^2 \right) }$
of the averaged CCF
$|\gamma_{i,j}|$.
As expected, the coherence length and the global degree of coherence decreased nearly linear with the exit slit opening. The coherence length in horizontal direction was determined to be
12 $\mu$m
for all exit slit openings (not shown here).

The characterization of the coherence properties of the X-rays was performed in order to use an optimal exit slit opening for ptychography. Although a
50 $\mu$m
exit slit opening provided the largest coherent fraction of the beam it also attenuated the beam strongly. Finally for ptychography measurements, the most coherent part of the beam provided by the
100 $\mu$m
exit slit with the modulus of the CCF
$|\gamma_{i,j}| > 0.8 $
was selected by the
2.6 $\mu$m
diameter pinhole
(see shaded area in Figure \ref{fig:coherenceCurves}(b) and (e)).

\section{Ptychography}

In far field ptychography the diffraction pattern amplitude
$A_i(q_x,q_y)$
in the detector plane is defined as the Fourier transform of the product of the probe
$P(x,y)$
and the projected object function
$O_i(x,y)$

\begin{equation}
A_i(q_x,q_y) = \mathcal{F} \{ P(x,y) \cdot O_i(x,y) \}
.
\label{eq:amplitude}
\end{equation}

Here
$x$ and $y$
are the coordinates in the sample plane,
$q_x$ and $q_y$
being momentum transfer values (i.e. the coordinates in reciprocal space) and
$\mathcal{F}$
denoting the Fourier transform operator. The index
$i$
spans along the individual scan position of the probe in the object plane. From the ptychographical reconstruction in transmission geometry
(see Figure \ref{fig:setup}(c))
one retrieves the complex object function
$O(x,y) = \exp(\mathrm{i} n k \Delta z(x,y))$
which depends on the wave number
$k = {2\pi} / \lambda$,
the object thickness
$\Delta z (x,y)$
and the complex refractive index
$n = 1 - \delta + \mathrm{i} \beta$.
Using the definition for
$n$
the object function can also be written as
\begin{equation}
O(x,y) = e^{ -k \beta \Delta z(x,y)} \cdot e^{ -\mathrm{i} k \delta \Delta z(x,y)}
.
\label{eq:objectFunction}
\end{equation}

Here the first term with the absorption coefficient
$\beta$
represents the projected amplitude and the exponent of the second term with the refraction coefficient
$\delta$
is the relative phase shift
$\Delta \varphi(x,y) = k \delta \Delta z(x,y)$.
Both terms contain the object specific response. First, we examined a strongly scattering tantalum (Ta) test object in the form of a Siemens star
(ATN/XRESO-50HC, NTT-AT, Japan).
The Siemens star was a lithographically manufactured sample with a smallest feature size of
50 nm.
Second, we measured a fossil diatom skeleton dispersed on a silicon nitride (Si$_3$N$_4$) membrane. With the
2.6 $\mu$m
pinhole the incident beam was shaped to a highly coherent and circular probe. Both samples were scanned on a rectangular scanning grid with a step size of
800 nm
in both
$x$ and $y$
direction perpendicular to the beam propagation axis. The detector was positioned in the far field at a distance of
58 cm
downstream from the sample.

\begin{figure}
\centering
\includegraphics[width=\linewidth]{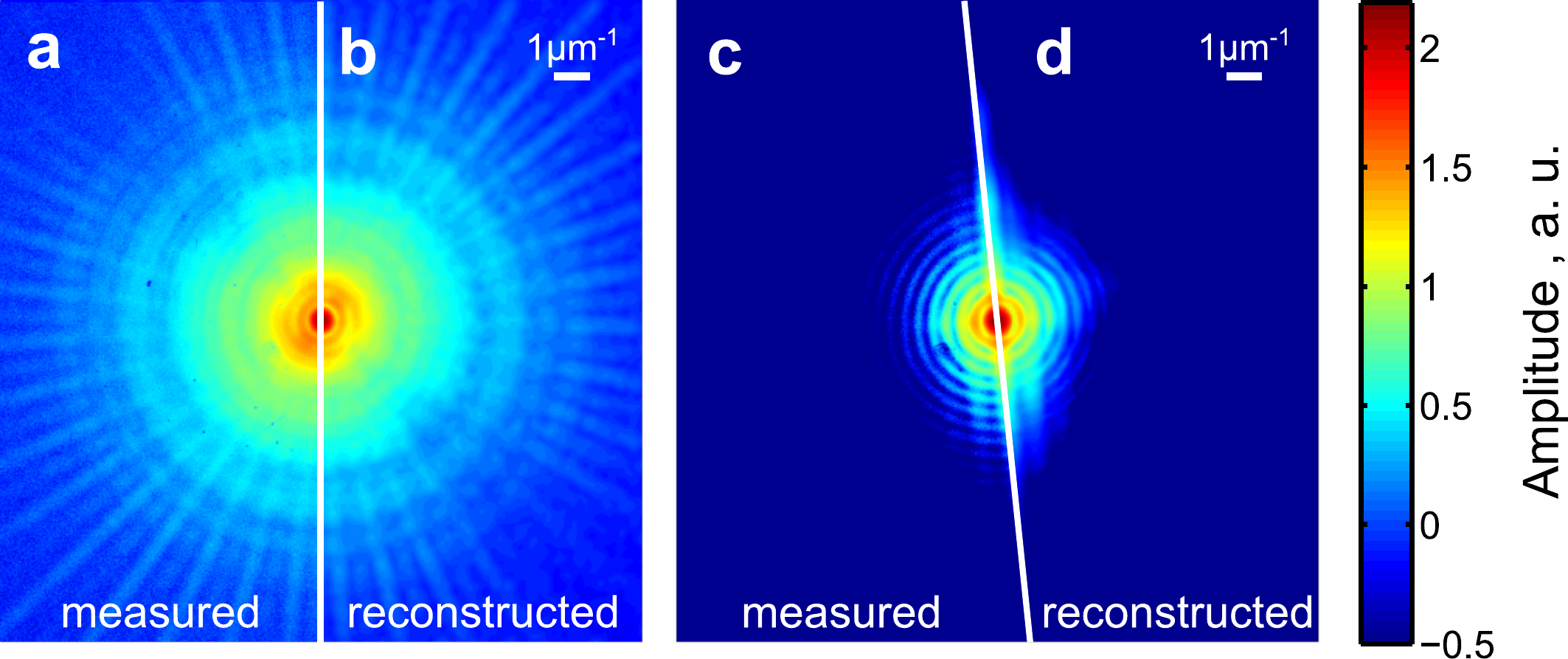}
\caption{Diffraction images averaged over all positions
$i$
for the Siemens star (a, b) and for the Diatom (c, d). Measured (a, c) and reconstructed diffraction patterns (b, d) are separated by the white line. The diffraction patterns are displayed on a logarithmic scale.}
\label{fig:diffractionPatterns}
\end{figure}

We implemented the ePIE algorithm
\cite{Maiden2009a}
to reconstruct both objects\footnote{Matlab (version 2013b) code was used for the entire analysis of the ptychography data. All calculations were executed on a graphics processing unit (GPU).}.
Initially the object function for all positions
$i$
was set to zero and the probe started with random values in amplitude and phase. The diffracted signal from the Siemens star was visible up to the edges of the detector. The reconstruction succeeded well over the whole diffraction plane
(see Figure \ref{fig:diffractionPatterns}(a, b)).
The diffraction signal from the diatom was comparably weak and reached the noise level already below the edges of the detector
(see Figure \ref{fig:diffractionPatterns}(c, d)).
As a consequence of the limited detector dynamic range, the signal above
7 $\mu$m$^{-1}$
was dominated by noise. To reduce the noise contribution in the diatom diffraction patterns we subtracted the total signal corresponding to the mean noise level of each pixel plus four times the standard deviation from this mean that was determined from ten dark images.

\subsection{Reconstruction of the probe function}

\begin{figure}
\centering
\includegraphics[width=\linewidth]{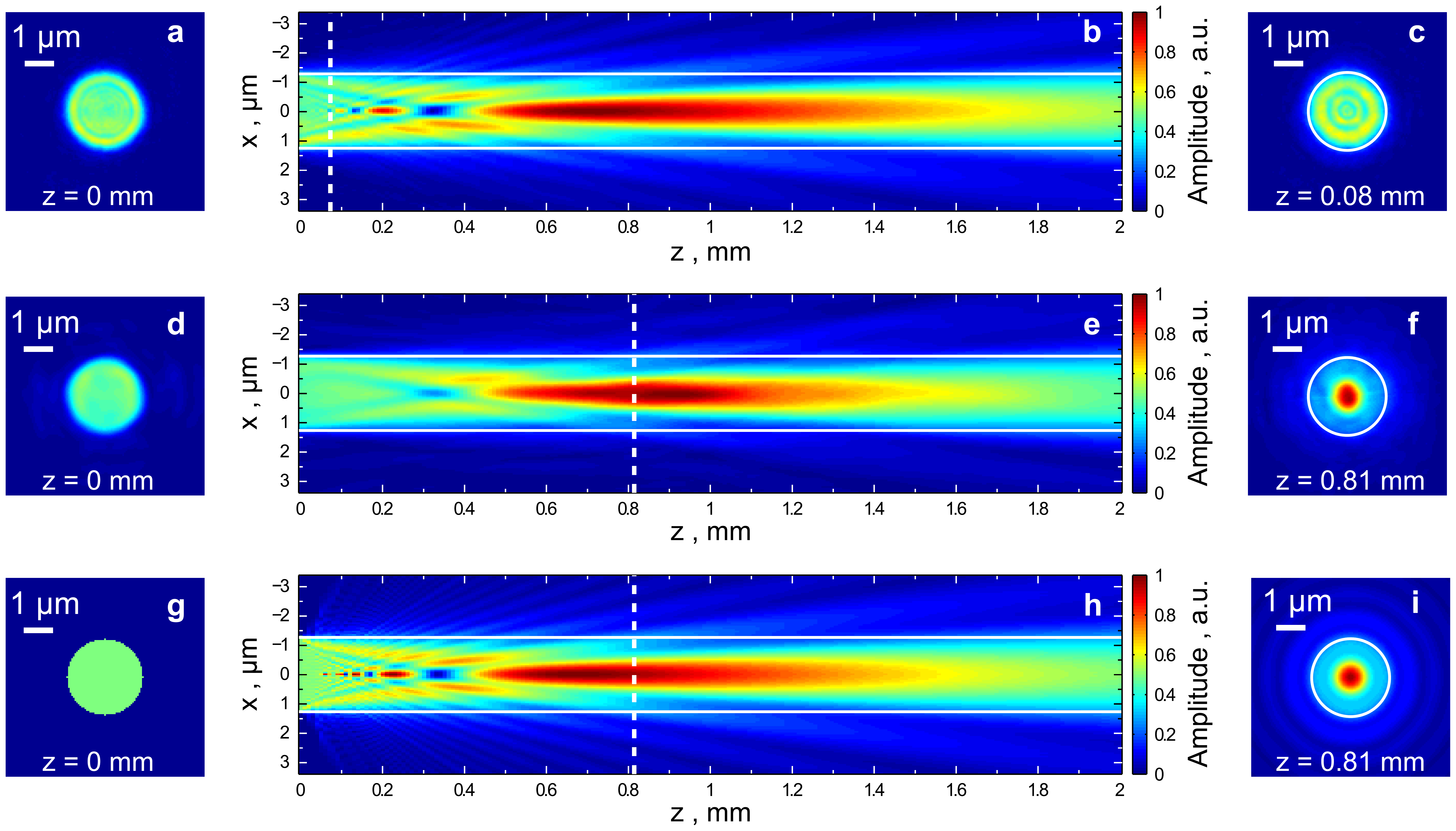}
\caption{Ptychographic probe function reconstruction from (a-c) the Siemens star and (d-e) the diatom. For comparison (g-i) an ideal probe originating as plane wave from the 2.6 $\mu$m pinhole. (a, d, g) Amplitude distribution at the position of the pinhole; (b, e, h) distribution of the wave field amplitude downstream the pinhole; (c, f, i) amplitude distribution of the probe amplitude at the position of the sample (Siemens star (c), or diatom (f, i). Dashed white lines indicate the position of the sample relative to the pinhole at $z = 0$. White lines indicate the pinhole size.}
\label{fig:probeFunctions}
\end{figure}

We exploited the strength of the ePIE algorithm and reconstructed the complex valued probe along with the object simultaneously. For both samples the probe was reconstructed and compared to a simulation in order to quantify the relative distance between the pinhole and the sample inside the HORST chamber. In the right column of
Figure \ref{fig:probeFunctions}
we show the reconstructed probes from the experiment together with a calculated ideal probe. The white lines indicate the pinhole size. The ideal probe wave field originated as a plane wave from the
2.6 $\mu$m
pinhole. In the middle column of
Figure \ref{fig:probeFunctions}
we show the amplitudes retrieved by forward and backward Fresnel propagation of the complex valued wave fields. In the left column we show the beam cross sections in the pinhole plane that result from the propagations. The probe incident on the Siemens star was determined to be  behind the pinhole at a distance of
0.08 mm.
It contained a fine structure visible as concentric fringes that is typical for near-field diffraction from a pinhole with a Fresnel number
8.5
(see Figure \ref{fig:probeFunctions}(c)).
We repeated the same procedure and found the diatom sample position at a distance of
0.81 mm
behind the pinhole
(see Figure \ref{fig:probeFunctions}(f)).
The propagation of the diatom probe wave field to the position at
$z = 0$
produced a well-shaped pinhole of
2.6 $\mu$m
diameter and practically constant amplitude across the pinhole.

We also compared our probe reconstructions with the simulation
(see Figure \ref{fig:probeFunctions}(g-i)).
The result of the simulation showed excellent agreement with the amplitude of the wave field distribution obtained from the Siemens star probe even at short distances behind the pinhole. The only difference was a slightly better contrast for the simulated propagation. A comparison with the wave field obtained from the diatom showed the same general behavior. However, it was lacking sharp features and high contrast which was both present in the Siemens star and simulated propagation. We attribute this effect to a lower signal produced by the diatom sample.

The comparison of the wave fields at the position of the diatom
($z=0.81$ mm)
showed similar amplitude distribution with a beam size of about
1 $\mu$m full width half maximum (FWHM).
Although this was considerably smaller than the pinhole (indicated as dashed-dotted black lines in
Figure \ref{fig:probeFunctions}(b, e, h))
the probe was still large enough to provide a sufficient overlap between adjacent illumination positions for ptychography.

One interesting result of this investigation is that one has to be careful if a pinhole is used in a ptychography experiment. The beam profile can be significantly different from the pinhole shape depending on the sample position behind that pinhole. In our experiment for example, at
$z=0.35$ mm
the beam amplitude was practically equal to zero in the center of the beam, at the same time at the position of
0.21 mm
behind the pinhole the beam had a sharp and narrow maximum with the size of
0.2 $\mu$m (FWHM)
as well as strong shoulders on both sides
(see Figure \ref{fig:probeFunctions}(b, h)).
Good conditions for our experiment were in principle the distances from
0.6 mm
to
1.6 mm
where the beam had a single pronounced peak and did not contain much amplitude in the wings outside of the geometrical pinhole region. A sample position
(smaller than 0.1 mm)
very close to the pinhole may also be beneficial because phase oscillations result in a structured probe that can improve the ptychographic reconstruction process in some cases
\cite{Maiden2013a,Quiney2005a}.

\subsection{Test pattern reconstruction}

\begin{figure}
\centering
\includegraphics[width=.65\linewidth]{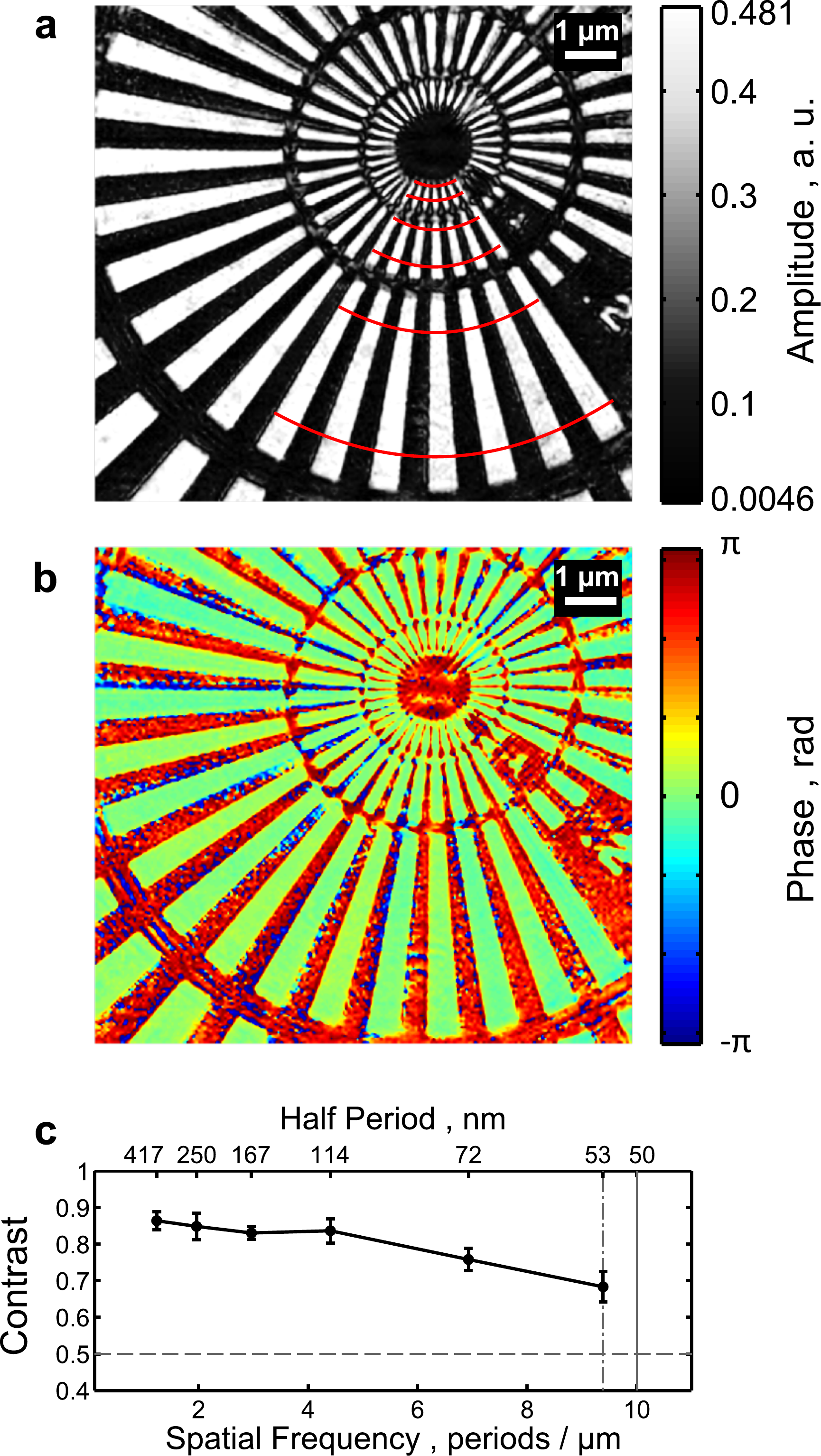}
\caption{Ptychographic reconstruction of the Siemens star test pattern. (a) Amplitude and (b) phase image. (c) Contrast
$C$
between high and low transmission as a function of spatial frequency from the angular scans denoted by red lines in the amplitude image.}
\label{fig:SimensStar}
\end{figure}

The dataset from the Siemens star consisted of
132
diffraction patterns
(12 horizontal x 11 vertical)
that covered an area of
10.4 x 9.6 $\mu$m$^2$
around the center of the test pattern. The reconstructed Siemens star amplitude is shown in
Figure \ref{fig:SimensStar}(a).
Black color denotes the strongly absorbing parts made of
270 nm
thick substrate layers from SiC, Si$_3$N$_4$, Ru including the test pattern with
500 nm
thick Ta with an expected transmission of
$T_{\mathrm{min}} = 4.6 \cdot 10^{-3}$.
In the white areas we expected a high transmission of
$T_{\mathrm{max}} = 0.48$
that was defined by the substrate only. These areas had zero phase variation, as can be seen in
Figure \ref{fig:SimensStar}(b).
In the low transmission parts a phase of
6.39 rad
was expected, however, undefined phases were observed. That could be explained by the phase value being close to
$2 \pi$
and small fluctuations causing complicated phase wrapping, both of which prevent a reliable and quantitative analysis
\cite{Giewekemeyer2011b}.

Performing a series of reconstructions we observed in some cases a wrong number of lines (other than 36) in angular direction while the radius of the concentric patterns did not deviate from the expectation. This effect was also observed previously
\cite{Burdet2014a}
and appears in case of not accurately known sample to detector distance.
With our detailed probe function analysis we finally determined an accurate sample to detector distance of
$578.5 \pm10$ $\mu$m and obtained the excellent reconstruction shown in Figure \ref{fig:SimensStar}.

We used multiple angular line scans to plot the reconstructed amplitude contrast
$C = \left( A_{\mathrm{max}} - A_{\mathrm{min}} \right) / \left( A_{\mathrm{max}} + A_{\mathrm{min}} \right)$
as a function of spatial frequency
(see Figure \ref{fig:SimensStar}(c)),
where
$A_{\mathrm{max}}$
and
$A_{\mathrm{min}}$
are the averaged maximum and minimum amplitudes of each line scan respectively. The error bars represent the standard deviation from the averaged contrast. The contrast decayed slowly from
0.87
at
1 period/$\mu$m
to
0.7
at the maximum spatial frequency of
9.4 periods/$\mu$m,
indicating a very good visibility for all spatial frequencies present in the reconstructed object. The maximum spatial frequency corresponds to a half period length of
53 nm
and was equal to the pixel size of the reconstruction. Since the diffracted signal was cut at the detector edges in reciprocal space
(see Figure \ref{fig:diffractionPatterns}(a, b))
the obtained resolution was limited only by the detector size.

\subsection{Fossil diatom reconstruction}

\begin{figure}
\centering
\includegraphics[width=.65\linewidth]{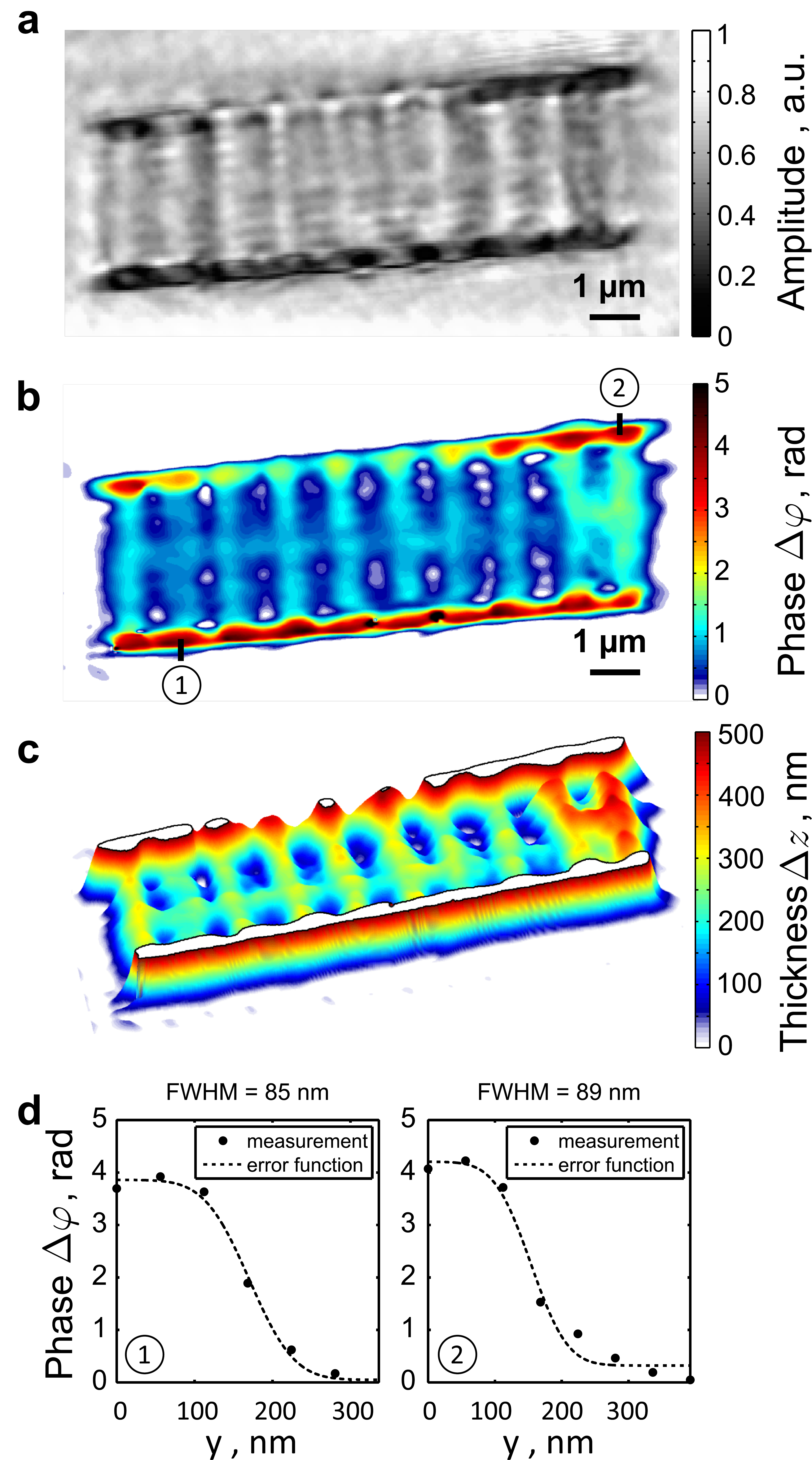}
\caption{Ptychographic reconstruction of the amplitude (a) and the phase (b) of the fossil diatom. (c) Thickness map and (d) FWHM values of two error function fits along the black lines indicated in the phase reconstruction in (b).}
\label{fig:Diatom}
\end{figure}

In the case of the fossil diatom the dataset consisted of
119
diffraction patterns
(17 horizontal x 7 vertical)
covering an area of
14.4 x 6.4 $\mu$µm$^2$.
As before in the case of the Siemens star reconstruction, we used our detailed probe function analysis to determine the accurate sample to detector distance and in this way avoided the problems of reconstruction of periodic structures. The reconstructed amplitude of the fossil diatom is shown in
Figure \ref{fig:Diatom}(a). The reconstructed and unwrapped phase of the sample is shown in
Figure \ref{fig:Diatom}(b).
The color scheme displays the relative phase shift map
$\Delta\varphi(x,y)$
between the substrate (shown in white) and the diatom.
Assuming uniform density of
SiO$_2$ (2.2 g/cm$^3$) we determined the refraction coefficient
$\delta$
at
500 eV photon energy to be
$\delta = 1.39 \cdot 10^{-3}$
\cite{Henke1993a}.
This allowed us to convert the relative phase shift
$\Delta\varphi(x,y)$
between the substrate and the diatom to a map of projected material thickness by applying the relation
$\Delta z(x,y) = \Delta\varphi(x,y) / (k \delta)$.
Diatoms of the \emph{Fragilariopsis cylindrus} species have a cylindrically curved shape
(see \cite{kang1992a}).
To avoid an ambiguous thickness profile we show the projected thickness in Figure
\ref{fig:Diatom}
up to a height of
500 nm.

Ten equidistantly spaced ribs to be
1 $\mu$m
period are visible in the reconstructed images. The length and width of each rib were estimated with
3 $\mu$m
and
250 nm
respectively. The fine structure that appeared in the form of a perforation and which is well pronounced in the amplitude image in Figure \ref{fig:Diatom}(a) had a period of
200 nm
in vertical direction. All the discovered features of the fossil diatom that we investigated here are specific for the species \emph{Fragilariopsis cylindrus} and are comparable with earlier studies \cite{kang1992a, Bertilson2009a}.

From line scans, which were extracted from the phase reconstruction, we determined the FWHM values of the fitted error functions.
(see Figure \ref{fig:Diatom}(d)).
Using the FWHM values, the resolution of our ptychographic diatom reconstruction was below
90 nm
and, consequently,
30$\,\%$
better than in a previously published paper
\cite{Giewekemeyer2011b}.

\section{Summary}

An experiment for high resolution ptychographical imaging of extended samples in the water window at
500 eV
with highly coherent X-rays from the P04 beamlime at PETRA III has been presented. The spatial coherence was characterized with an efficient NRA method and high coherence of the X-ray beam was demonstrated. The global degree of coherence in the vertical direction varied from $10\%$ to $73\%$ depending on the setting of the exit slit of the monochromator. For ptychographic measurements an optimal exit slit size of 100 $\mu$m was used providing a coherence length of
$4.1$ $\mu$m
and a degree of coherence of
$35\%$
at the sample position. With these settings the most coherent part of the beam was selected by the
2 $\mu$m
pinhole that produced high contrast diffraction patterns in our ptychographic experiment.

We obtained important knowledge about the fine features of the probe function by the propagation analysis that allowed us to determine the sample position inside of the HORST vacuum chamber with high precision. This turned out to be important in imaging periodic samples to avoid the artifacts caused by an inaccurately known sample to detector distance. The ptychographic reconstruction of the Siemens star test pattern was obtained up to
53 nm
resolution and was limited only by the detector size. In the case of the weakly scattering fossil diatom we obtained the amplitude and phase of the transmission function quantitatively with a resolution better than
90 nm.
Finally, we exploited the phase reconstruction and obtained a quantitative thickness map with details that allowed a comparison with the diatom species reported in the literature.

Ptychography in the water window with its sensitivity for oxygen and carbon contrast is especially promising to image hydrated biological cells. With the cryo-extension of HORST vacuum chamber its capability to image extended biological samples on the nanoscale will become feasible.
We are also planning to extend our research to three-dimensional (3D) ptychography with high spatial resolution
\cite{Dierolf2010a}.
This will allow us to determine the fine structure of diatoms on the nanoscale in 3D.


\ack{Acknowledgements}

The authors acknowledge the financial support from the BMBF 05K10VH4 and the Virtual Institute VH-VI 403. The support and fruitful discussions with E. Weckert are greatly acknowledged. The authors are thankful to A. Singer and S. Lazarev for fruitful discussions and to A. Schropp for a careful reading of the manuscript.

\referencelist[bib] 


\end{document}